\documentclass{emulateapj}

\newcommand{\Te} {$T_{\rm eff}$}
\newcommand{\logg} {$\log g$}

\begin{document}

\title{Detailed compositional analysis of the heavily polluted DBZ
  white dwarf SDSS~J073842.56+183509.06: A window on planet
  formation?\footnote{Some of the data presented herein were obtained
    at the W.M.  Keck Observatory and the MMT Observatory. Keck is
    operated as a scientific partnership among the California
    Institute of Technology, the University of California and the
    National Aeronautics and Space Administration. The Observatory was
    made possible by the generous financial support of the W.M. Keck
    Foundation. The MMT is a joint facility of the Smithsonian
    Institution and the University of Arizona.}}

\author{P. Dufour\altaffilmark{1}, M. Kilic\altaffilmark{2},
  G. Fontaine\altaffilmark{1}, P. Bergeron\altaffilmark{1},
  C. Melis\altaffilmark{3}, J. Bochanski\altaffilmark{4}}
  \email{dufourpa@astro.umontreal.ca}

\altaffiltext{1}{D\'epartement de Physique, Universit\'e de
  Montr\'eal, Montr\'eal, QC H3C 3J7, Canada}

\altaffiltext{2}{Homer L. Dodge Department of Physics and Astronomy,
  University of Oklahoma, 440 W. Brooks St., Norman, OK, 73019, USA}

\altaffiltext{3}{Center for Astrophysics and Space Sciences,
  University of California, San Diego, CA 92093-0424, USA}

\altaffiltext{4}{Department of Astronomy and Astrophysics, The
  Pennsylvania State University, University Park, PA 16802, USA }

\begin{abstract}

  We present a new model atmosphere analysis of the most metal
  contaminated white dwarf known, the DBZ
  SDSS~J073842.56+183509.06. Using new high resolution spectroscopic
  observations taken with Keck and Magellan, we determine precise
  atmospheric parameters and measure abundances of 14 elements heavier
  than helium. We also report new {\em Spitzer} mid-infrared
  photometric data that are used to better constrain the properties of
  the debris disk orbiting this star. Our detailed analysis, which
  combines data taken from 7 different observational facilities
  (GALEX, Gemini, Keck, Magellan, MMT, SDSS and Spitzer) clearly
  demonstrate that J0738+1835 is accreting large amounts of rocky
  terrestrial-like material that has been tidally disrupted into a
  debris disk. We estimate that the body responsible for the
  photospheric metal contamination was at least as large Ceres, but
  was much drier, with less than 1\% of the mass contained in the form
  of water ice, indicating that it formed interior to the snow line
  around its parent star. We also find a correlation between the
  abundances (relative to Mg and bulk Earth) and the condensation
  temperature; refractory species are clearly depleted while the more
  volatile elements are possibly enhanced. This could be the signature
  of a body that formed in a lower temperature environment than where
  Earth formed. Alternatively, we could be witnessing the remains of a
  differentiated body that lost a large part of its outer layers.

\end{abstract}


\section{INTRODUCTION}

The heavy elements observed in the spectra of otherwise pure hydrogen
or helium atmosphere white dwarfs represent handy fingerprints of
asteroids/rocky planets that survived the late phases of stellar
evolution \citep{Zuckerman07, Klein2010,Klein2011,Dufour10}. There is
growing evidence that perhaps all metal polluted white dwarfs of
spectral type DZ, DBZ and DAZ have acquired their heavy material from
an orbiting debris disk reservoir whose origin is explained by the
tidal disruption of one or many large rocky bodies that ventured too
close to the star
\citep{debes,jura03,jura06,jura07,jura08,farihi10,farihi10b,Melis10}. Such
disks, which are easily detectable at infrared wavelengths, are now
discovered with an accelerating pace with more than 20 cases uncovered
in the last 6 years alone \citep[see][and references
therein]{farihi09,Farihi11,kilic11}.

Meanwhile, high resolution observations in the optical also revealed
the presence of weak calcium lines in many white dwarfs, suggesting
that at least 25\% of their progenitors possessed orbiting planetary
bodies with masses often comparable to that of the largest solar
system asteroids \citep{Zuckerman03,Zuckerman10}. Unfortunately, the
majority of the known metal polluted white dwarfs only show one or two
heavy elements in the optical, hardly enough to make detailed
planetary composition inquiries \citep{dufour07}. In a few extreme
cases, however, a handful of elements are observed, opening a unique
window for studying rocky exoplanet compositions. The first
investigations, those of GD~362 and GD~40 \citep{Zuckerman07,
  Klein2010} revealed striking similarity between the accreted bodies
and bulk Earth material, providing the first comprehensive
measurements of the bulk elemental composition of ancient planetary
systems. Analysis of similar objects, based on Keck high resolution
observations, quickly followed these pionering studies, bringing to 6
the number of metal polluted white dwarfs with more than 8 heavy
elements detected in the optical \citep[the previous two along with
G241$-$6, NLTT~43806, PG~1225$-$079 and
HS~2253+8023,][]{Zuckerman10,Zuckerman11,Klein2011}.

Although the global abundance patterns found for these few objects
were relatively well explained by the accretion of rocky planetary
material similar to bulk Earth, closer looks reveal a wide range of
compositional diversity from star to star (or, equivalently, planets
to planets), especially for the most refractory
elements. Interestingly, dynamical simulations such as those of
\citet{Bond, Bond2} also predict a wide variety of chemical properties
and structure compared to what is observed in the solar system. The
most metal polluted white dwarfs could thus represent perfect testbeds
to discriminate between various planet formation scenarios, allowing
the first empirical verification of extrasolar terrestrial planet
formation simulations. However, other processes such as post AGB
thermal heating, collisions and wind stripping, could also play an
important role in the final bulk composition that is measured at a
white dwarf photosphere. Understanding the physical mechanisms
responsible for the various observed elemental abundance patterns
represents one of the biggest challenge of this relatively young field
of research. As the sample of well studied metal contaminated white
dwarfs increases, a more comprehensive picture should emerge, leading
to a better understanding of how, both individually and statistically,
these planetary systems form and evolve.

Recently, \citet{Dufour10} discovered an extremely polluted DBZ white
dwarf, SDSS J073842.56+183509.06 (hereafter J0738+1835) in the Sloan
Digital Sky Survey. A preliminary analysis, based on medium resolution
optical spectroscopy, revealed the presence of O, Mg, Si, Ca and Fe in
record quantities. Given that Keck/HIRES spectroscopy of the heavily
polluted white dwarf GD 362 revealed 15 different heavy elements
\citep{Zuckerman07}, three of which are detected in low resolution
spectroscopy \citep{gianninas}, we anticipated that higher resolution
observations of J0738+1835 would reveal several additional elements
not detected in \citet{Dufour10}'s low resolution data.

This paper reports a detailed analysis of J0738+1835 based on new high
resolution optical spectra taken at the Keck and Magellan
telescopes. We also report new mid-infrared photometric data taken with
{\em Spitzer} which, when combined with Gemini $JHK$ photometry, provides
better constraints on the debris disk physical properties. In addition
to the heavy elements already detected by \citet{Dufour10}, our new
spectroscopic observations of J0738+1835 reveal the presence of 9 new
elements (Na, Al, Sc, Ti, V, Cr, Mn, Co and Ni), for a total of 14
chemical species heavier than helium. This is the second highest
number of heavy elements ever observed at a white dwarf photosphere
\citep[the champion, with 15 different metals, being
GD~362;][]{Zuckerman07} thus making J0738+1835 a new and important
member of the very select group of polluted white dwarfs with eight or
more elements heavier than helium detected optically \citep{Zuckerman11}.

In \S~\ref{observation}, we describe the new observations. Our detailed
analysis of the spectroscopic and photometric observations follows in
\S~\ref{analysis} while the results are discussed and summarized in
\S~\ref{results} and \S~\ref{conclusion}.

\section{OBSERVATIONS}\label{observation}

Our detailed study of SDSS J0738+1835 uses data taken from 7 different
observational facilities (GALEX, SDSS, Gemini, MMT, Keck, Magellan and
Spitzer), with a strong emphasis on new Keck HIRES spectroscopy. Part
of these data have already been presented in the preliminary analysis
of \citet{Dufour10} and are not discussed further here. Below we
describe the new data.

\subsection{Keck/HIRES}

We used the High Resolution Echelle Spectrometer
\citep[HIRES,][]{Vogt} on the Keck I 10m telescope (Mauna Kea
Observatory) to obtain high resolution optical spectroscopy of
J0738+1835 over 4 hours on UT 2011 March 23. We use the blue
collimator and the 1.148$\arcsec$ slit with the cross disperser angle
of 1.075, providing wavelength coverage from 3100 \AA\ to 5990 \AA\
with a resolving power of 37,000. The seeing started at 0.9" with an
average around 1.1 $\arcsec$. We use the MAuna Kea Echelle Extraction
(MAKEE) software for data analysis. We flat-field the exposures,
optimally extract the spectra from the two dimensional image traced by
the standard star Feige 34 observed on the same night, remove cosmic
rays, and wavelength calibrate the spectra using Th-Ar lamps. MAKEE
does not extract spectra in a few of the partially covered orders. We
use the HIRES Redux package written by J. X. Prochaska to extract
these missing orders from the MAKEE pipeline. The HIRES observations
of J0738+1835 is thus composed of 58 short spectroscopic segments,
typically 60 to 90 \AA\ wide, with some wavelength overlap in adjacent
orders. The overlapping parts have been co-added to slightly improve
the signal-to-noise ratio.

\subsection{Magellan/MagE Optical Spectroscopy}

Moderate-resolution optical spectra of J0738+1835 were obtained on
2011 March 19 (UT) with the Magellan Echellette
\citep[MagE;][]{2008SPIE.7014E.169M}, mounted on the 6.5m Landon Clay
Telescope at Las Campanas Observatory.  Conditions during the
observations were clear with 0.6$\arcsec$ seeing.  Two exposures of
1200~s each were obtained over an airmass range of 1.48--1.49 using
the 0.7$\arcsec$ slit aligned with the parallactic angle; this setup
provided 3200--10050~\AA\ spectroscopy at a resolution $\approx$ 8000
(although only the red part longward of 6000 \AA, where some important
oxygen lines are present, is used in the present analysis since the
Keck observations covers the rest with a better resolution and
signal-to-noise ratio).

We also observed the spectrophotometric flux standard
Hiltner~600 \citep{1994PASP..106..566H} on the same night for flux
calibration. Th-Ar lamps were obtained after each source observation
for wavelength calibration. Order tracing and pixel response were
calibrated with internal Xe lamp flats and twilight sky flats,
respectively, obtained at the beginning of the night.  Data were
reduced using the MASE reduction pipeline \citep{2009PASP..121.1409B},
following standard procedures for order tracing, flat field
correction, vacuum wavelength calibration (including heliocentric
correction), optimal source extraction, order stitching, and flux
calibration.

\subsection{MMT spectroscopy}

We used the 6.5m MMT equipped with the Blue Channel spectrograph to
obtain medium resolution spectroscopy of J0738+1835 over 109 minutes
on UT 2010 March 19.  We operate the spectrograph with the 832 line
mm$^{-1}$ grating in second order, providing wavelength coverage from
3170 \AA\ to 4100 \AA\ and a spectral resolution of $\approx$ 1
\AA. All spectra were obtained at the parallactic angle. We used
He-Ne-Ar comparison lamp exposures and blue spectrophotometric
standards \citep{Massey} for wavelength and flux calibration,
respectively. We reduce the data using standard IRAF routines.

\subsection{Spitzer}

We used the warm {\em Spitzer} equipped with the InfraRed Array Camera
\citep[IRAC,][]{Fazio} to obtain mid-infrared photometry of J0738+1835
on UT 2010 December 1 as part of program 70023. We obtained 3.6
and 4.5 $\mu$m images with integration times of 100 seconds for nine
dither positions. We use the IDL astrolib packages to perform aperture
photometry on the individual corrected Basic Calibrated Data frames
from the latest available pipeline reduction. We get similar results
using 2 and 3 pixel apertures, but we quote the results using the
smallest aperture since it has the smallest errors.

Following the IRAC calibration procedures, we correct for the location
of the source in the array before averaging the fluxes of each of the
dithered frames at each wavelength.  We also correct the Channel 1
(3.6 $\mu$m) photometry for the pixel-phase-dependence. We estimate
the photometric error bars from the observed scatter in the nine
images corresponding to the dither positions. We add the 3\% absolute
calibration error in quadrature \citep{reach05}. \citet{reach09}
demonstrate that the color corrections for dusty white dwarfs like
G29-38 are small (0.4-0.5\%) for channels 1 and 2. We ignore these
corrections for J0738+1835. 

The final observed flux and uncertainties at 3.6 $\mu$m and 4.5 $\mu$m
are 81.1 $\pm$ 2.7 $\mu$Jy and 76.7 $\pm$ 2.7 $\mu$Jy
respectively.

\section{DETAILED ANALYSIS}\label{analysis}

\subsection{Model Atmospheres and Fitting Technique}

The model atmosphere and synthetic spectrum calculations are performed
using the same code that was used in the first attempt to model
J0738+1835 \citep{Dufour10} with the exception that we now use atomic
data provided by the Vienna Atomic Line Database
(VALD\footnote{http://vald.astro.univie.ac.at/$\sim$vald/php/vald.php})
instead of the Kurucz line list.

As a first step, we calculated a model structure using the same
parameters (\Te, \logg\ and abundances) as \citet{Dufour10}, including
all the transitions (from 200 to 10,000 \AA) present in our
linelist. Using this thermodynamic structure, we next calculated
several grids of synthetic spectra, one for each element of interest,
keeping the abundance of all other elements fixed to their original
values. For example, the Fe grid covers a range of abundance from log
[n(Fe)/n(He)] = $-$4.0 to $-$6.0 in steps of 0.5 dex with the abundances
of all other metals fixed to those previously determined. The various
grids typically vary by log [n(Z)/n(He)] = $\pm$1.5 dex (in step of
0.5 dex) around the previously determined \citep[or expected by
assuming $\rm CI$ chondrites ratios from][]{lodders} abundance. These
grids are then used to determine the different elemental abundances as
follows.

Since the optical spectra for J0738+1835 are extremely crowded with
several hundred lines, many of which overlap, it is not practical to
fit each line individually using equivalent width
measurements. Instead, we made short segments, typically 15 to 30
Angstrom wide, centered on each of the strongest absorption features
observed in our dataset (i.e., those that are unambiguously detected
and deeper than $\sim$10\% of the continuum flux level). Then, each
segment (a few hundreds) are fitted separately with the appropriate
grids.

\begin{figure*}[!ht]
  \plotone{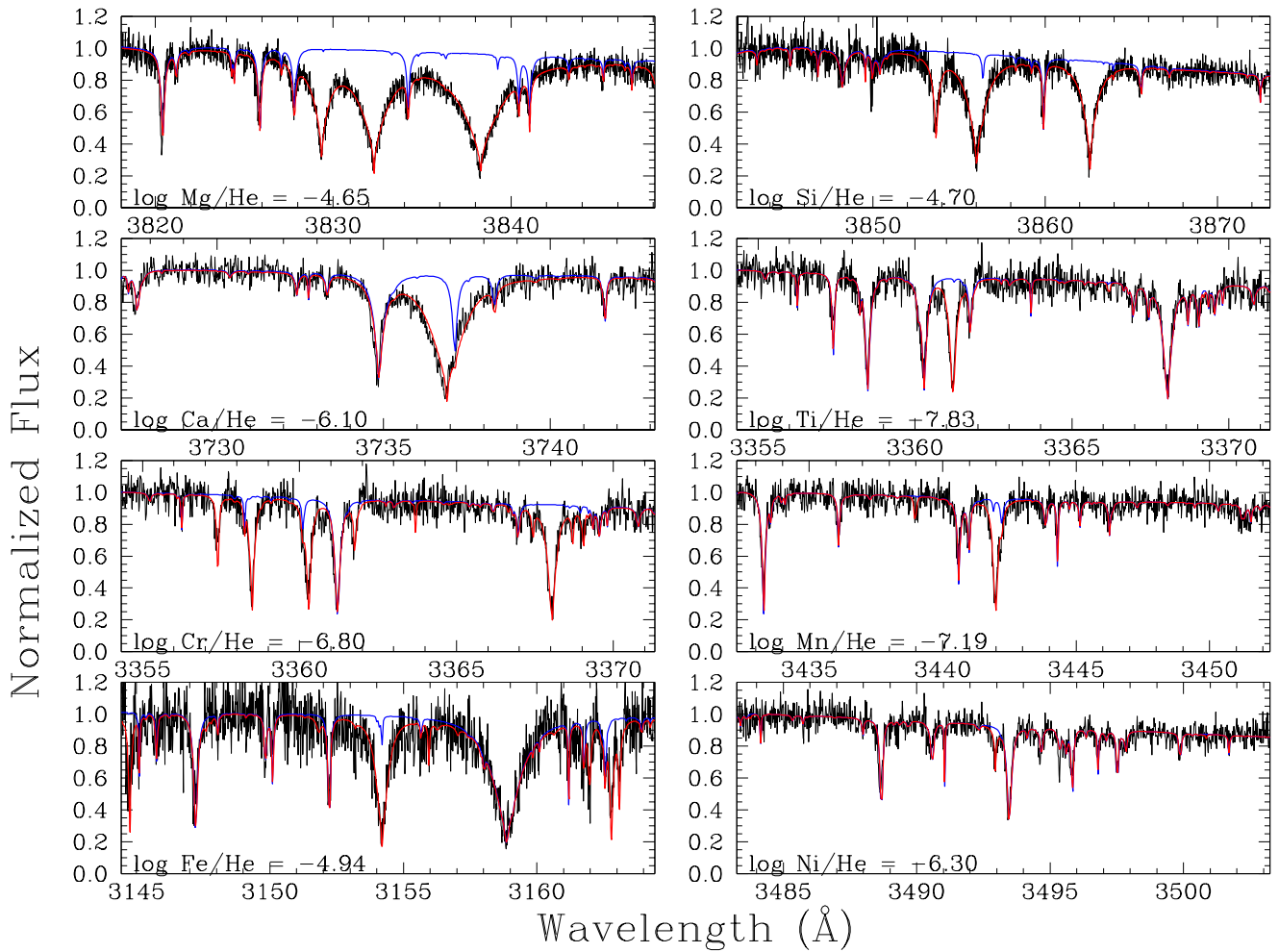} \figcaption{Typical example of fits to small HIRES
segments. Only one element at a time is fitted in each panel, all
other elements being kept fixed to the value determined in the
previous iteration (see text). The determined abundance of the element
of interest, which is not the final abundance reported in Table 2, is
indicated in the lower left corner of each panel. The red line is our
best fit while the blue line is a synthetic spectrum where the element
of interest has been removed to help localize the regions where the
fitted lines contribute. The observations have been artificially
shifted to air wavelengths in order to facilitate line identification
using line lists.}
\end{figure*}

Our fits are done by minimizing the value of the $\chi^2$ taken as the
sum of the difference between the observed and model fluxes over the
wavelength range of interest, all frequency points being given an
equal weight. We consider the abundance, the solid angle and a linear
term to account for the slope (the numerous HIRES orders are not flux
calibrated) as free parameters. This method allows a very good
determination of the local continuum and naturally takes into account
the contribution of the absorption from other elements. It is thus
particularly well suited for situations where it is difficult to
precisely determine equivalent widths of individual lines, either
because line blending is important or the signal-to-noise ratio is
marginal. We show in Figure 1 typical examples of segment fits. In
order to better visualize the fitted features, we plotted in blue a
synthetic spectrum in which the element that is fitted has completely
been removed. We note that in many cases, more than one line from a
given fitted element can be found in a single segment. In these cases,
the determined abundance of the segment is essentially a weighted
average of the lines, more weight (more frequency points) being put on
the strongest lines. We finally take as the measured abundance of an
element the average over all the segments.

It must be noted, however, that these abundances were determined from
synthetic spectra calculated with a structure corresponding to the
\citet{Dufour10} solution. The presence of new species, as well as the
slightly different abundances that we now derive (and, to a smaller
extent, the use of the VALD data), has a small impact on the
thermodynamic structure of our model. Hence, in order to obtain a
self-consistent solution, it is necessary to repeat the above
mentioned procedure (i.e., recalculate all the grids and redo the
fits) but, this time, starting from a model structure calculated with
the newly determined set of abundances. We iterate until the input
abundances in the model structure and the final measurements from our
grids are in agreement.

However, even this solution can not be considered final since such an
analysis is done for a fixed value of \Te\ and \logg\ (13,600~K and
8.5, respectively). These parameters were originally determined from
fitting the HeI line profiles as well as simultaneously accommodating
the MgI and MgII line strengths \citep[see][]{Dufour10}. As discussed
in \citet{Klein2010,Klein2011}, the absolute abundance of the various
elements can vary significantly (up to 0.4 dex in some stars) with
small variations of \Te and \logg. The relative composition of the
polluting elements, however, are much less sensitive to the exact
final parameters adopted and can thus be used as a high precision
measurement of the accreted planets/planetesimals (see
below). Nevertheless, we believe it is best to take full advantage of
the opportunities provided by our dataset, and proceed in obtaining the
best atmospheric parameters as possible by reevaluating the effective
temperature, the surface gravity and the elemental abundances in a
self consistent way.

We thus repeated the whole analysis as described above but, this time,
starting with model structures calculated at effective temperatures of
13,300 and 13,900~K \citep[corresponding to the estimated
uncertainties of $\pm$300~K in][]{Dufour10}. We then focus on elements
for which lines from two different ionization state are present. In
particular, we select a subsample of the strongest well isolated lines
of Mg and Fe, the elements that show the largest number of lines from
both the neutral and ionized state. In Figures 2 and 3, we plot the
average abundance for each ion as a function of effective temperature,
taking the standard deviation of the set of individual abundances as
error bars. Clearly, a much better agreement between the two
ionization states is obtained for a slightly higher effective
temperature than first obtained by \citet{Dufour10}. A remarkable
agreement between the various ion abundances is reached for an
effective temperature of $\sim$ 13,950 K. This effective temperature
estimate, which is determined solely from Mg and Fe lines, is
completely independent of the uncertain treatment of van der Waals HeI
line broadening.

\begin{figure}[!ht]
  \plotone{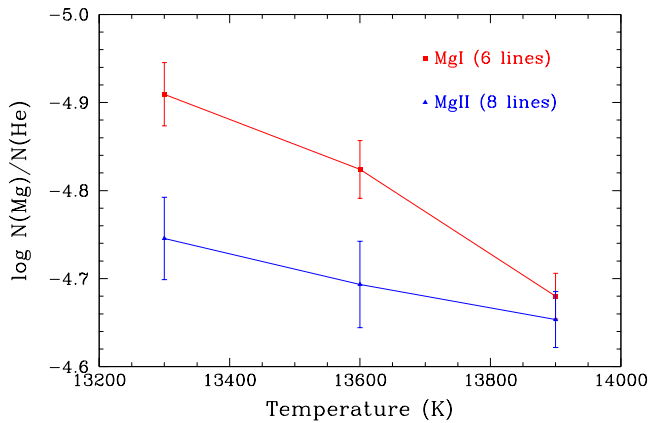} \figcaption[] {Average abundances of magnesium lines as a
  function of effective temperature for two different ionization
  states. The error bars represent the standard deviation of the set
  of individual. A much better internal coherence is found for an
  effective temperature near 13,950~K.}
\end{figure}

\begin{figure}[!ht]
  \plotone{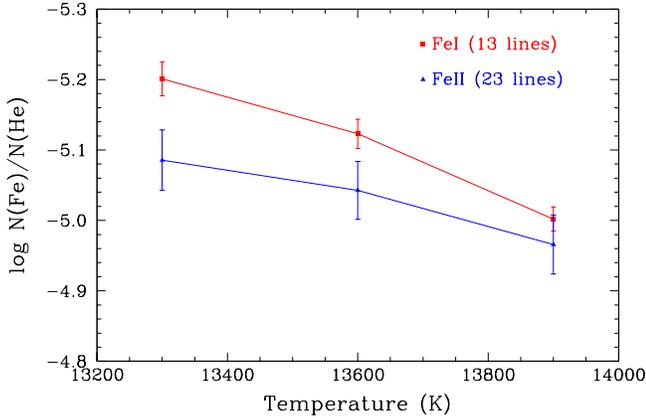} \figcaption[] {Same as Figure 2, but for iron.}
\end{figure}

Next, we take this newly determined effective temperature and repeat
again the above procedure but this time with surface gravities between
\logg = 8.3 and 8.7. Our results, summarized in Figure 4 and 5,
indicate that a surface gravity lower than \logg = 8.5 better
reproduces the different ion abundances. A surface gravity of
\logg$\sim$ 8.4 is preferred by the Mg lines while a slightly lower
surface gravity is required for the Fe lines, although a value of
\logg$\sim$8.4 is compatible with the error bars.

\begin{figure}[!ht]
  \plotone{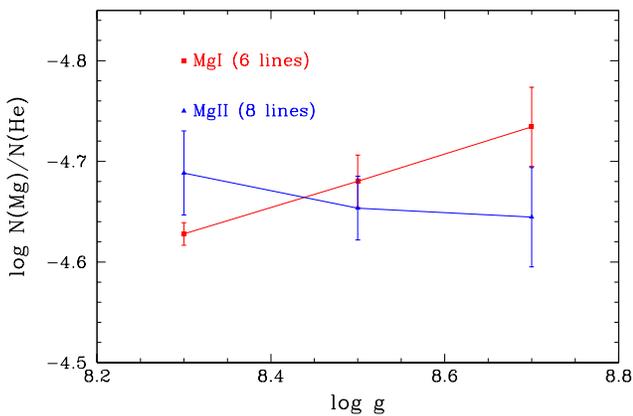} \figcaption[] {Average abundances of magnesium lines
    as a function of surface gravity for two different ionization
    states. The error bars represent the standard deviation of the set
    of individual. A slightly lower surface gravity than \logg = 8.5
    appears to be preferable.}
\end{figure}

\begin{figure}[!ht]
  \plotone{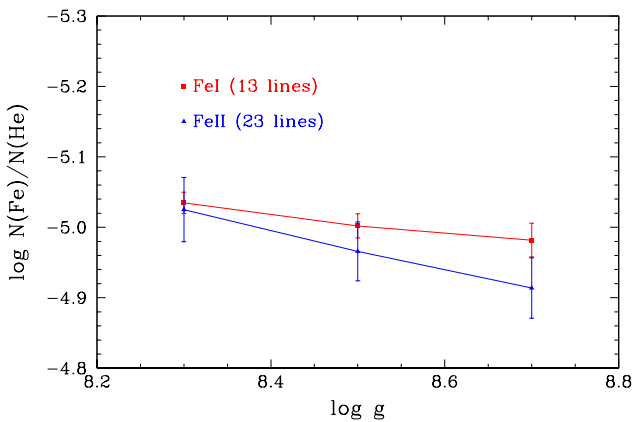} \figcaption[] {Same as Figure 4, but for iron.}
\end{figure}

We thus adopt, as our final parameters, an effective temperature of
13,950~K and a surface gravity of \logg $\approx$ 8.4 with
conservative uncertainties of $\pm$100~K and $\pm$0.2 dex
respectively\footnote{Such a level of accuracy for the effective
  temperature ($\sim$0.7\%) is among the most precise ever determined
  for a white dwarf star.}. These uncertainties are next properly
propagated for the determination of errors on the cooling age,
luminosity, mass, radius and distance of the star (obtained using the
solid angle from $ugriz$ photometry). The evolutionary models used are
similar to those described in \citet{FBB01} but with C/O cores, and
thickness of the helium and hydrogen layers of respectively q(He) =
10$^{-2}$ and q(H) = 10$^{-10}$, which are representative of
helium-rich atmosphere white dwarfs.  To evaluate the abundance
uncertainties, we proceed in a similar manner to that described in
\citet{Klein2010} and vary the model \Te~ and \logg~ by their
uncertainties, one at a time, determine the average change in
abundances for each parameter, and add them in quadrature to the
standard deviation of the set of individual abundances. Our final
parameters are presented in Table~1 and 2 while our best fit models
are plotted over the spectroscopic data for the most interesting HIRES
and MagE orders in Figures 6 to 12.

\begin{deluxetable}{lccccc}
\tabletypesize{\scriptsize}
\tablecolumns{11}
\tablewidth{0pt}
\tablecaption{Stellar parameters adopted for SDSS J0738+1835}
\tablehead{Parameter & Value}
\startdata
$T_{\rm eff}$(K)  & 13950 $\pm$ 100    \\
$\log g$         & 8.4 $\pm$ 0.2    \\
$M_{\rm WD}/M_{\odot}$     & 0.841  $\pm$ 0.131  \\
$M_{\rm init}/M_{\odot}$  &   4.47 $\pm$0.36 $^a$  \\
$R/R_{\odot}$     & 0.00958 $\pm$ 0.0015  \\
$\log L/L_{\odot}$ & $-$2.50 $\pm$0.12 \\
$D$              & 147 pc $\pm$23 \\
Cooling Age      & 477 Myr $\pm$ 160\\
$\log (M_{\rm He}/M_{\star})$ &  $-$6.41 $+0.56$/$-0.28$ \\
\enddata
\tablenotetext{a}{Initial mass of the main sequence progenitor
  calculated using the Initial-Final Mass Relation of
  \citet{williams09}}
\end{deluxetable}

\begin{deluxetable*}{lccccc}
\tabletypesize{\scriptsize}
\tablecolumns{11}
\tablewidth{0pt}
\tablecaption{}
\tablehead{Element & log $[n(Z)/n(He)]_{phot}$  & $M_{\rm CVZ}/(10^{21}$g) & log $\tau_{set}$(yr) & $[n(Z)/n(Fe)]_{acc}$ & $\dot M /(10^8g~s^{-1})$ }
\startdata

1  ~H & -5.73 $\pm$ 0.17 & 0.310  & $\infty$ &  \ldots & \ldots \\
8  ~O & -3.81 $\pm$ 0.19 & 407.86 & 5.244 & 9.52 & 740.2 \\
11 Na & -6.36 $\pm$ 0.16 & 1.639  & 5.238 & $2.7\times 10^{-2}$ & 3.02 \\
12 Mg & -4.68 $\pm$ 0.07 & 83.33 & 5.258 & 1.24 & 146.38 \\
13 Al & -6.39 $\pm$ 0.11 & 1.792 & 5.244 & $2.5\times 10^{-2}$  & 3.25\\
14 Si & -4.90 $\pm$ 0.16 & 57.99 & 5.248 & 0.77 & 104.36 \\
20 Ca & -6.23 $\pm$ 0.15 & 3.907 & 5.044 & $5.8\times 10^{-2}$ & 11.24 \\
21 Sc & -9.55 $\pm$ 0.18 & 2.05$\times 10^{-3}$ & 5.010 & $2.9\times 10^{-5}$& 6.38$\times 10^{-3}$ \\
22 Ti & -7.95 $\pm$ 0.11 & 8.87$\times 10^{-2}$ & 5.007 & $1.2\times 10^{-3}$  & 0.278 \\
23 V  & -8.50 $\pm$ 0.17 & 2.65$\times 10^{-2}$ & 5.006 & $3.4\times 10^{-4}$ & 8.31$\times 10^{-2}$\\
24 Cr & -6.76 $\pm$ 0.12 & 1.492 & 5.026 & $1.8\times 10^{-2}$ & 4.48 \\
25 Mn & -7.11 $\pm$ 0.11 & 0.693 & 5.028 & $7.7\times 10^{-3}$ & 2.07 \\
26 Fe & -4.98 $\pm$ 0.09 & 94.91& 5.047 & 1.00 & 271.32 \\
27 Co & -7.76 $\pm$ 0.19 & 0.165 & 5.042 & $1.7\times 10^{-3}$ & 0.479 \\
28 Ni & -6.31 $\pm$ 0.10 & 4.721 & 5.063 & $4.6\times 10^{-2}$ & 12.997 \\
\hline
\hline
 &           &          &  &  &  \\
Total &           & 658.95   &  &  & 1301.7 
\enddata
\end{deluxetable*}

\begin{figure*}[!ht]
  \plotone{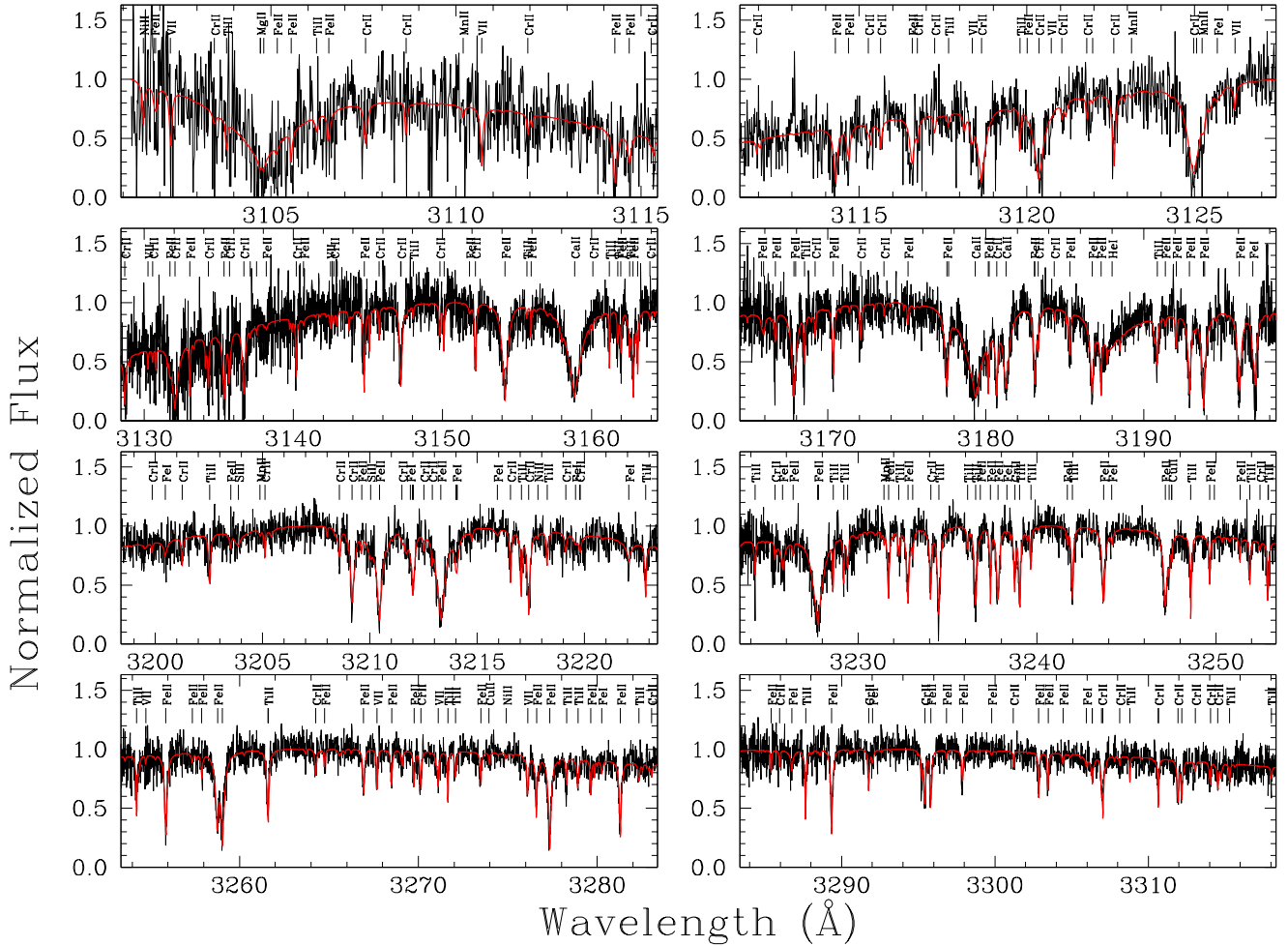} \figcaption[] {Display of our final solution (in
    red) over the most interesting HIRES orders. Ticks and labels
    identify the strongest features present in the spectra. The
    observations have been artificially shifted to air wavelengths in
    order to facilitate line identification using available line
    lists.}
\end{figure*}

\begin{figure*}[!ht]
  \plotone{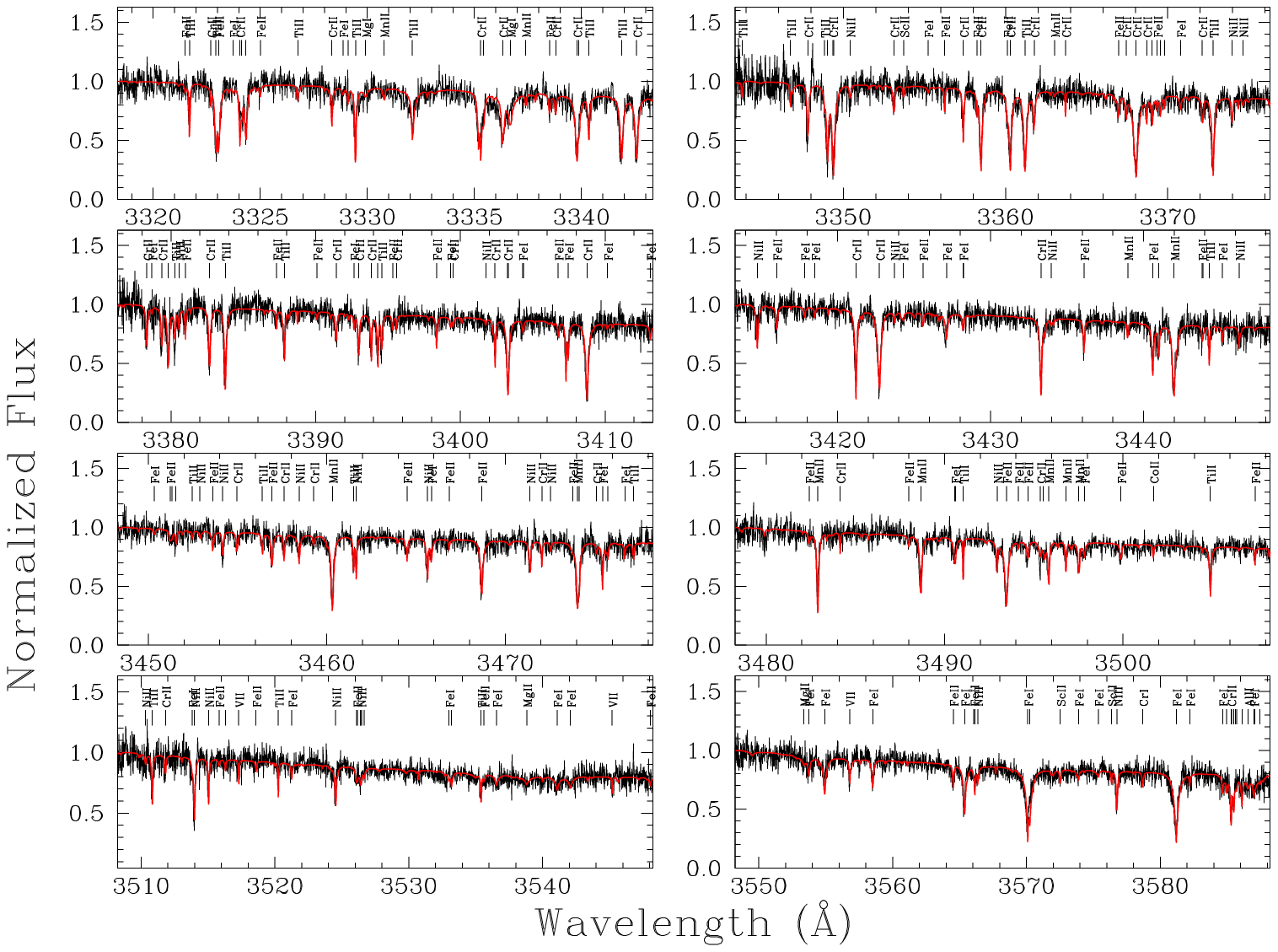} \figcaption[] {Same as previous Figure.}
\end{figure*}

\begin{figure*}[!ht]
  \plotone{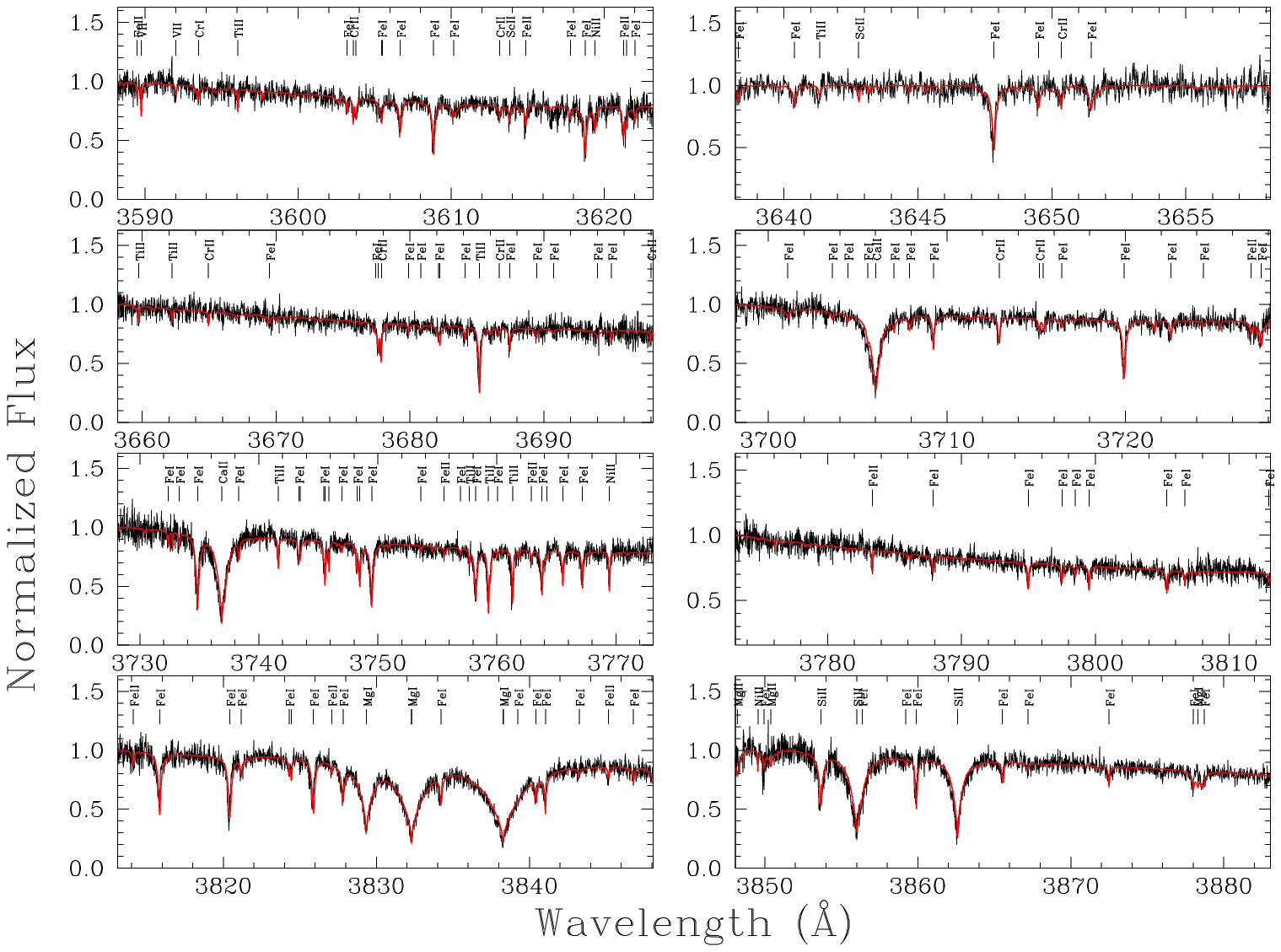} \figcaption[] {Same as previous Figure.}
\end{figure*}

\begin{figure*}[!ht]
  \plotone{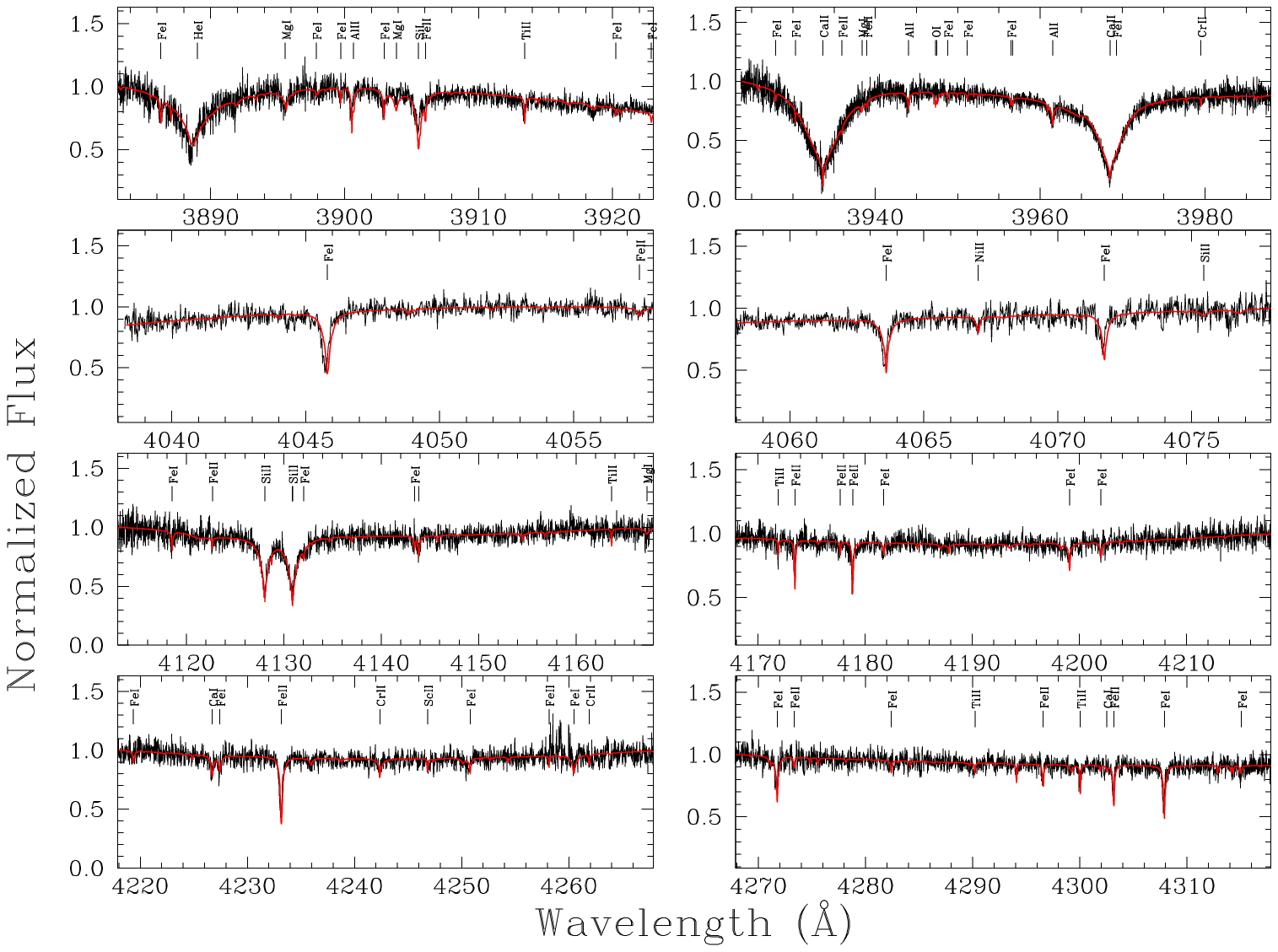} \figcaption[] {Same as previous Figure.}
\end{figure*}

\begin{figure*}[!ht]
  \plotone{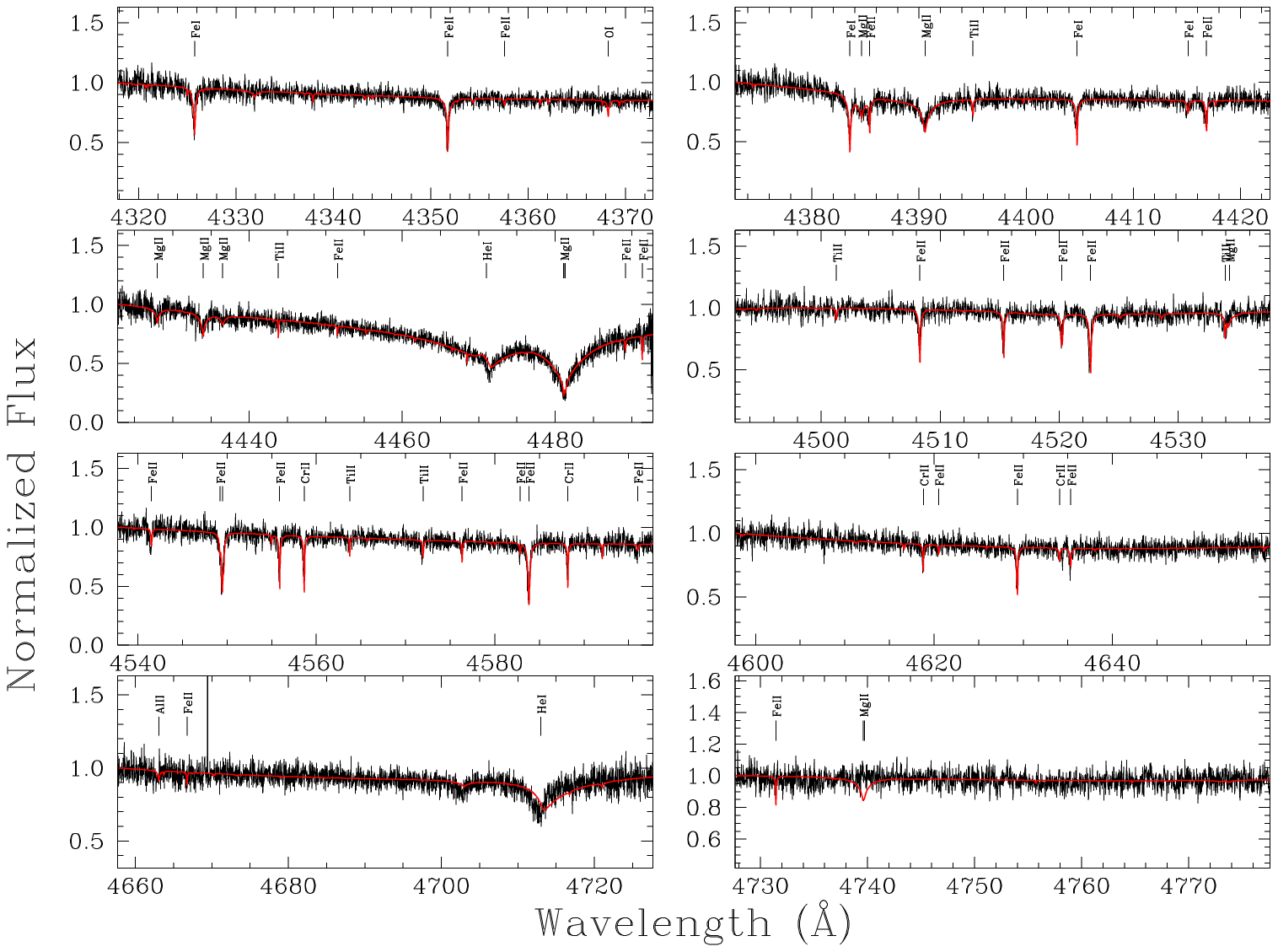} \figcaption[] {Same as previous Figure.}
\end{figure*}

\begin{figure*}[!ht]
  \plotone{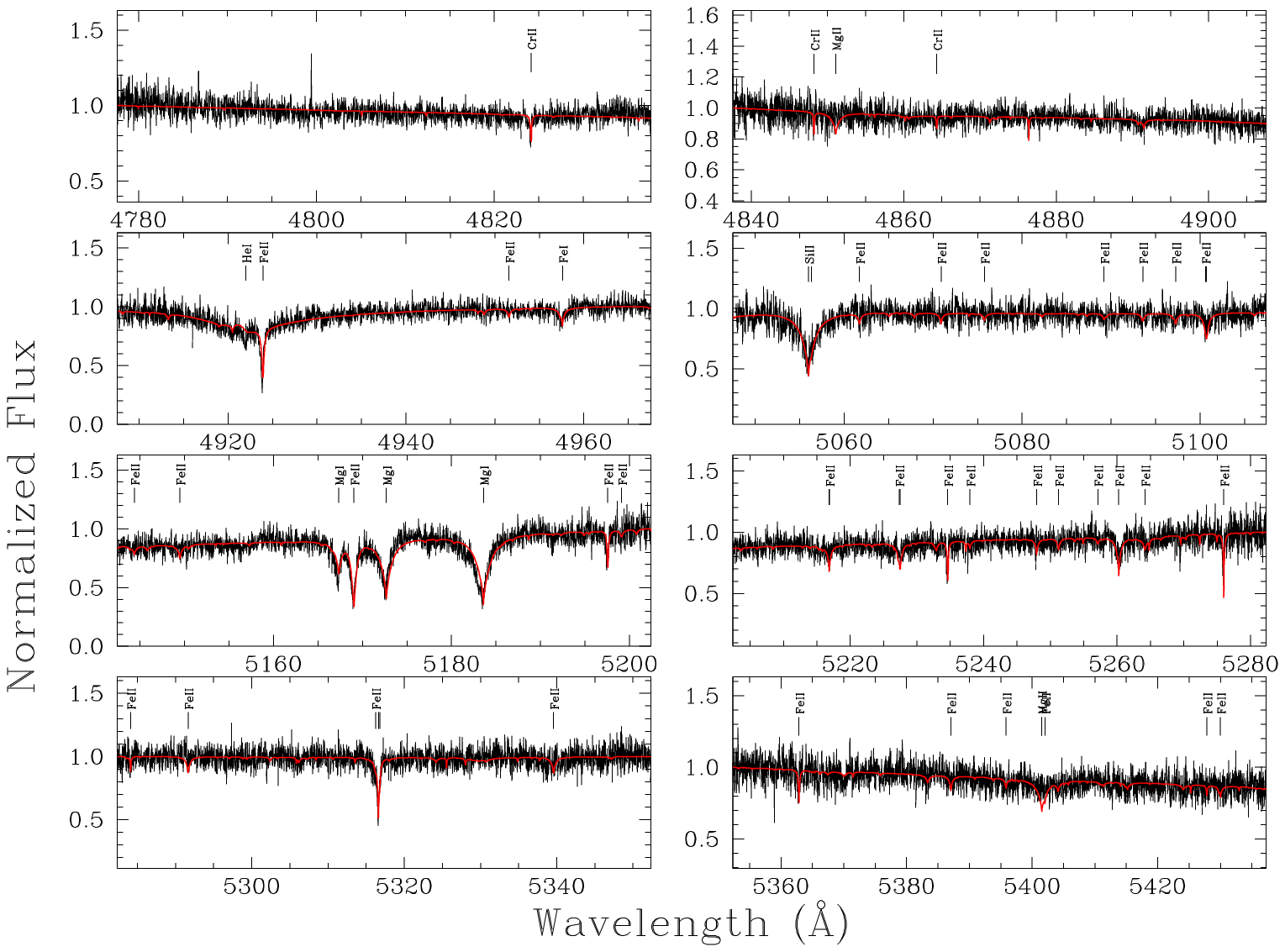} \figcaption[] {Same as previous Figure.}
\end{figure*}

\begin{figure*}[!ht]
  \plotone{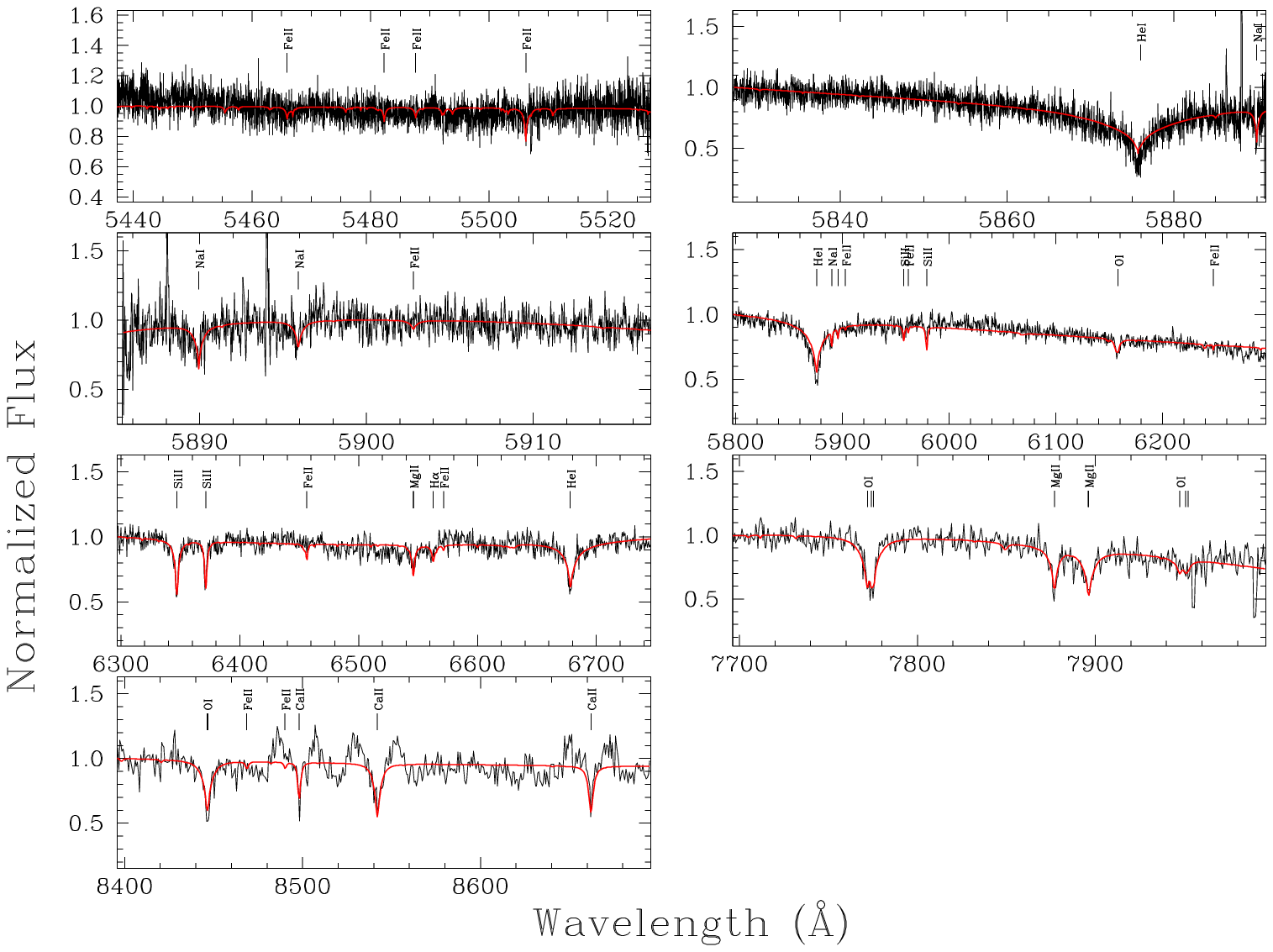} \figcaption[] {Same as previous Figure. The last
    five panels display MagE spectroscopic data. The double-peaked gas
    disk emission features of CaII first reported by
    \citet{Gaensicke11} are evident in the last panel.}
\end{figure*}

We also, a posteriori, verified explicitly that this new solution was
still compatible with the lower resolution MMT spectroscopic data used
in \citet[][]{Dufour10}. Figure 13 shows that the slight change in
atmospheric parameters that we derive here did not have any noticeable
impact on the quality of the fit to the various HeI lines in our "old"
MMT data. In fact, we even observe that most of the discrepancies in
the predicted line strengths of Fe lines \citep[see][Figure
3]{Dufour10} are now almost completely removed. We attribute this
result on our improvement in the atmospheric parameters as well as the
use of better atomic line data (VALD).

\begin{figure}[!ht]
  \plotone{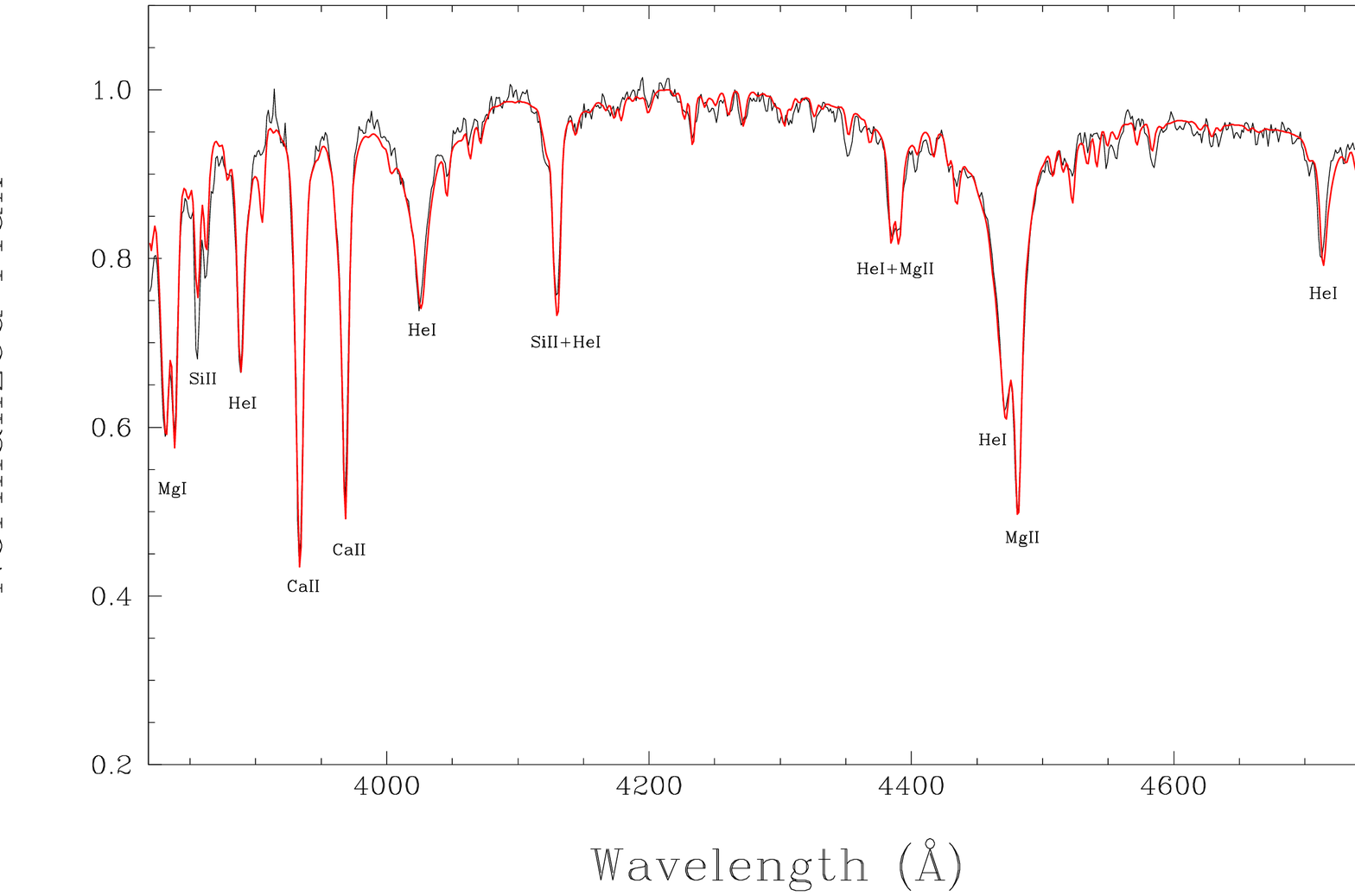} \figcaption[] {Display of our final solution (in
    red) over the lower resolution MMT data used in \citet[][see their
    Figure 3]{Dufour10}. The strongest lines are identified while all
    the other numerous non-identified lines are from iron. The
    discrepancies in Fe line strengths noted by \citet{Dufour10} have
    almost all disappear (see text).}
\end{figure}

It is interesting to note that while the HeI line profiles seem well
reproduced by our models when observed at relatively low resolution,
such is not the case at higher resolution. Indeed, detailed inspection
of strong HeI line at $\lambda\lambda$ 3889, 4471, 4713, 4922 and 5876
indicates that the synthetic models poorly reproduce the HIRES
observations (the MagE observations of HeI $\lambda\lambda$ 5876 and
6678, taken at lower resolution, are more satisfactory). Similar
problems in HeI line profiles have also been observed in other cool
DBZ white dwarfs \citep[see, in particular, the narrow absorption core
component in HeI $\lambda$5876 for GD~362, GD~16 and
PG~1225$-$079,][]{Zuckerman07,Klein2011}. Note that varying the
atmospheric parameters (\Te/\logg) from the optimal values determined
above does not improve our fit to the helium lines; the discrepancy
appear to reside in the predicted line shape. The explanation for the
features morphology eludes us at this time. Until this issue is
resolved, we believe it is safer to use the ionization balance of
heavy elements to obtain the effective temperature and surface gravity
with high precision

We also present in Figure 14 a similar exercise using unpublished
$\approx$1 \AA\ resolution MMT spectroscopic observations that covers
the bluer part of the electromagnetic spectrum. While the
signal-to-noise ratio and resolution of these observations were first
judged to be insufficient to provide significant improvements over
\citet[][]{Dufour10}'s solution, they are nevertheless useful to
confirm that no significant change in the elemental abundances of the
major species has occurred between the $\sim$1 year that separates the
MMT and Keck observing runs. This is not surprising given that this
object is most probably accreting material at a steady state and that
the diffusion timescales of the heavy elements are of the order of
$10^5$ years (see below).

\begin{figure}[!ht]
  \plotone{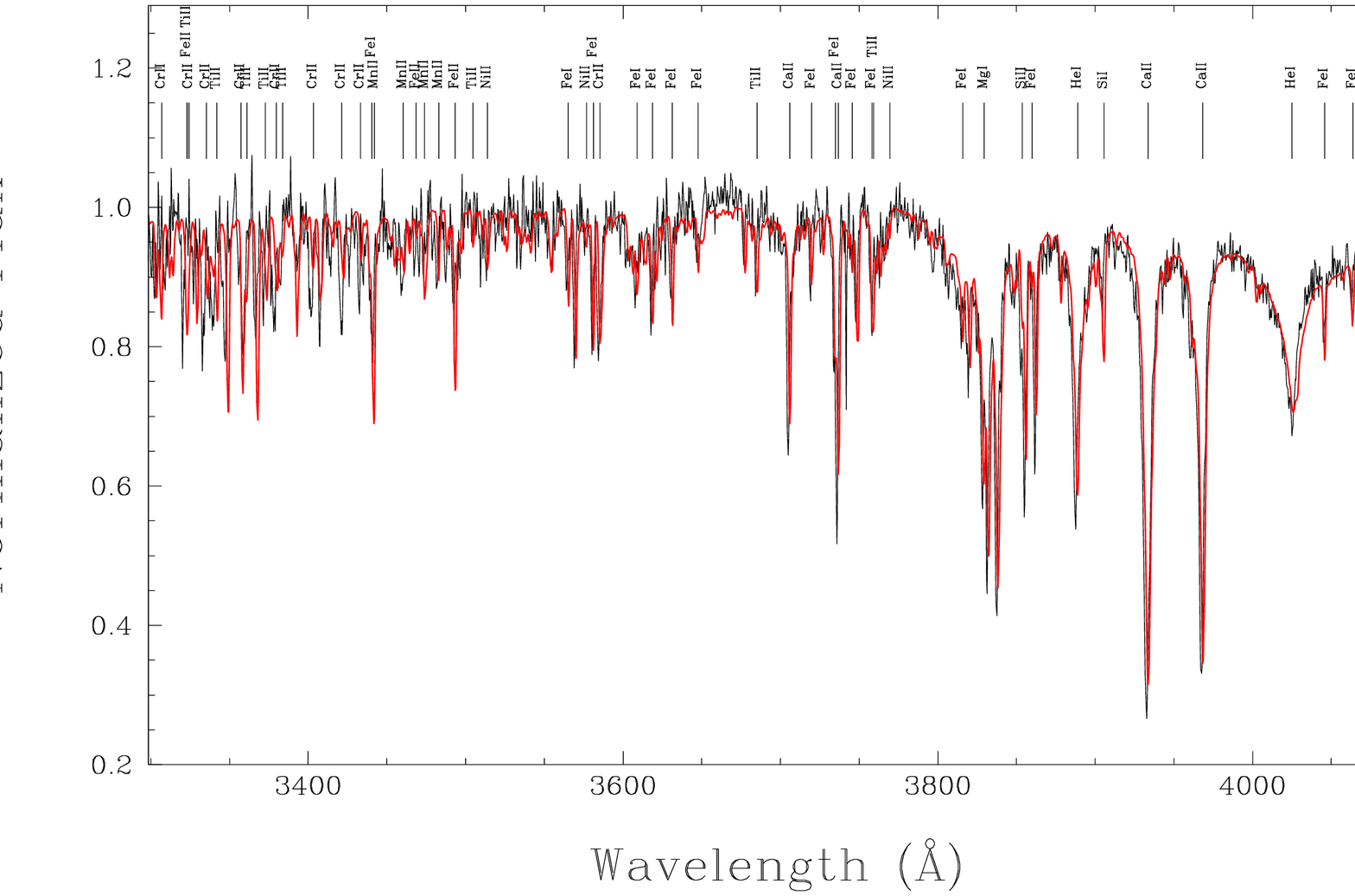} \figcaption[] {Display of our final solution (in
    red) over new unpublished medium resolution MMT data. Ticks and
    labels identify the strongest features present in the spectra.}
\end{figure}

Although the final abundance determination has been obtained only from
an average of a subsample of all the lines present in the Keck and
Magellan spectroscopic datasets, we find that the synthetic spectrum
calculated with our final solution agrees remarkably well with the
observations for the vast majority of the lines over the whole
spectrum (see Figures 6 to 12). This can most probably be attributed to
the fact that the method described above has yielded a very accurate set
of atmospheric parameters that minimized the dispersion among the set
of individual measurements. Indeed, we find that the standard
deviation of the set of individual measurements for a given element
vary typically between 0.04 to 0.10 dex (with a few exceptions,
notably oxygen, that have slightly larger dispersions; see below).

There are, however, a few individual lines which significantly depart
from our final optimal solution (most probably due to discrepancies in
atomic line data or broadening parameters). These can be observed in
Figures 6 to 12 as features that are not very well reproduced by the
final synthetic spectrum (see, for example, lines near
$\lambda\lambda$4370, 4740, 4850, 5400, 5505). Overall, fortunately,
the final atmospheric parameters are not significantly affected by
these lines since their contribution tend, for the few included in the
final set, to be averaged out from using multiple lines from each
elements (O, Na, Al, Sc and Co, which rely on 3 or less individual
measurements, must thus be considered more uncertain on that
account). 

We note, finally, that our observations for J0738+1835 show no sign of
emission cores in the CaII H\&K lines, as observed for PG~1225$-$079
\citep{Klein2011}. \citet{Klein2011} proposed that NLTE effects could
possibly cause such emission since these line cores are formed high up
in the atmosphere. However, comparing the thermodynamic structures of
PG~1225$-$079, calculated with \citet{Klein2011}'s parameters, with
that of J0738+1835 argues against this interpretation (not to mention
that NLTE effect are not expected in such a low effective temperature
white dwarf). Indeed, we find that the atmospheric pressure in
J0738+1835 is lower than that of PG~1225$-$079, while its gas
temperature is higher at every depth point in the atmosphere. Since
both of these should favor NLTE effects (less collisions and a more
intense radiation field), if NLTE was the explanation, J0738+1835 (and
other similar DBZ white dwarfs) would be even more likely to show
cores in emission. Since they are not observed in any other object,
PG~1225$-$079 is more likely a unique and interesting case of white
dwarf chromospheric activity.

It is interesting to note, however, that our MagE spectroscopic
observations clearly show evidence of the IR CaII double-peaked
emissions (see Figure 12). These features, which were already observed
in the SDSS spectroscopic data \citep[see][]{Gaensicke11}, are the
typical signature of gaseous debris disks rotating around a white
dwarf \citep{Gaensicke06,Gaensicke07,Gaensicke08}.

It has been proposed that gas disk-hosting systems could be the result
of multiple infalling bodies originating from a solar system-like
asteroid belt \citep{jura08}. However, the total mass of heavy element
presently found in J0738+1835's convective zone represent a very large
fraction of the mass of an asteroid belt. While it is possible to
explain this large amount by the accretion of several small asteroids,
as newly-shredded asteroids dust arrives near the pre-existing disk,
destructive grain-grain collisions should rapidely cause the disk to
evolve toward a large gaseous system \citep{jura08}. Given that a
significant dusty disk component is revealed from a large infrared
excess (see below), we are most likely withnessing a case where one
large object generated the dense dust disk which dominate the system
and that the small amount of gas observed is the result of much
smaller bodies hitting the dust disk from time to time. It is also
possible that we are watching the system in a transition phase where
the dusty disk is starting to slowly dissipate, through internal
motions, into a gaseous phase. Alternatively, the emission line flux
could be understood, as proposed by \citet{Gaensicke11}, by some
heating of the top layers of the disk with ultraviolet photons from
the white dwarf \citep[see also][]{Melis10}. Unfortunately, our
understanding of white dwarfs debris/gaseous disk is insufficient at
this point to speculate furthermore on the formation/evolution of
these components.

\subsection{Infrared Excess and Disk Model}

The presence of an accretion disk around J0738+1835 was unambiguously
established from $JHK$ excess photometry in \citet{Dufour10}. However,
due to the lack of mid-infrared data, the outer temperature was not
well constrained. Here we use {\em Spitzer} (IRAC) 3.6 and 4.5 $\mu$m
data combined with Gemini JHK photometry to get better constraints on
the physical parameters of the disk.

In the same manner as in \citet{Dufour10}, we use the optically thick
flat-disk models of \citet{jura03} and determine the best fit disk
parameters. We obtain $T_{in} = 1600 \pm 100$~K, $T_{out} = 900
^{+100}_{-200}$~K and an inclination angle of 40 $\pm$ 20 degree (see
Figure 15). The uncertainties for these parameters were obtained
following the method described in \citet{kilic11}. Briefly, a Monte
Carlo analysis is performed where the observed photometric fluxes f
are replaced with f + g$\delta$f, where $\delta$f is the error in flux
and g is a Gaussian deviate with zero mean and unit variance. For each
of 10,000 sets of modified photometry, the analysis is repeated to
derive new best-fit parameters for the disk. The interquartile range
of these parameters is adopted as the uncertainty. The larger low
range uncertainty found for the outer temperature is the result of not
having data beyond 5 microns since it is possible that the disk may
extend to larger radii with a corresponding cooler lower-limit for the
temperature. With $T_{out} = 900$~K, corresponding to an outer radii
of $\sim$ 0.1 R$_{\odot}$, the disk is well placed inside the tidal
radius of the white dwarf ($\sim$ 1 R$_{\odot}$), further
strengthening the idea that the origin of the disk, as well as the
material polluting the white dwarf's convection zone, is the tidal
disruption of a large rocky body.

\begin{figure}[!ht]
  \plotone{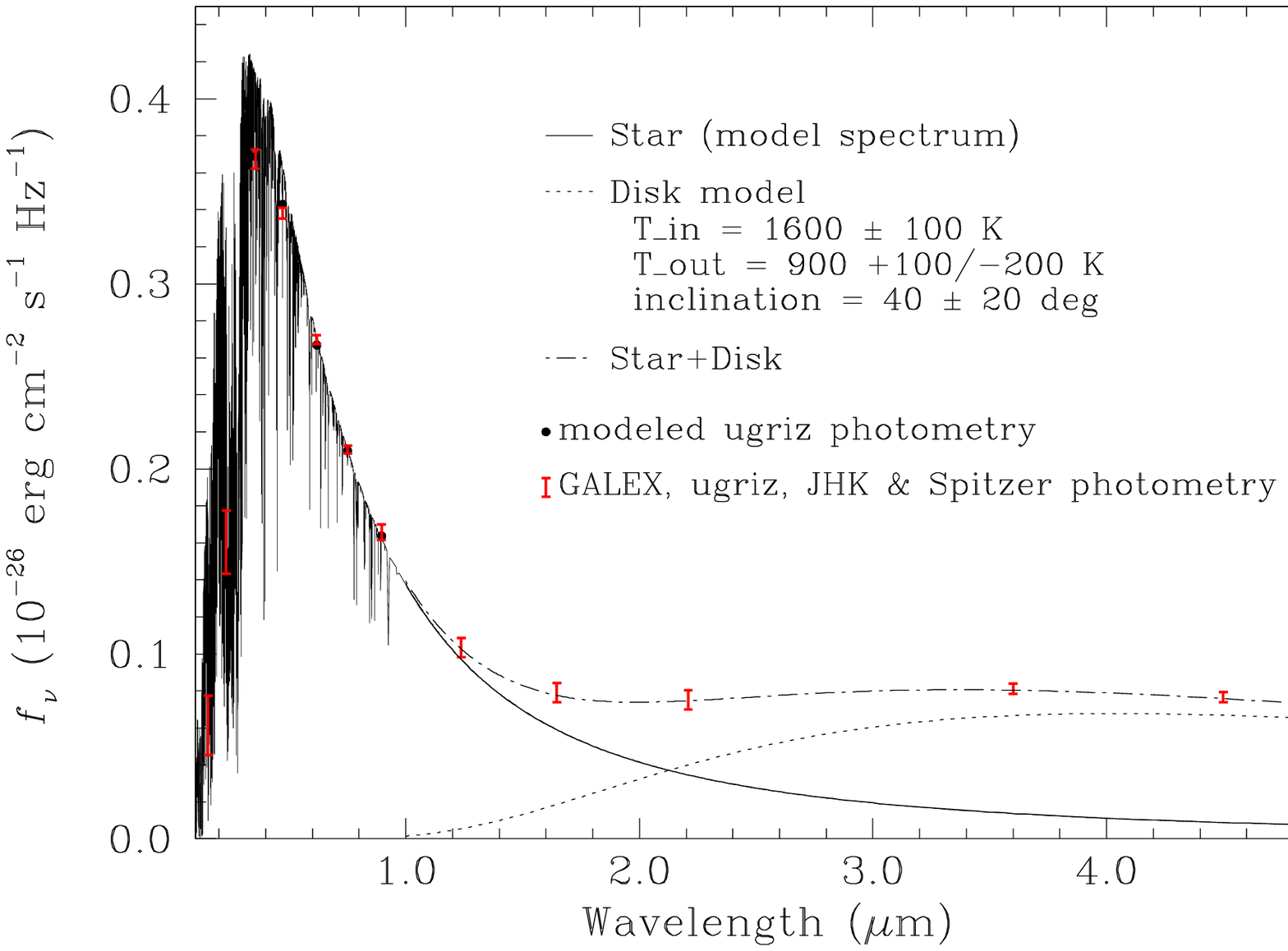} \figcaption[] {Photometric measurements in GALEX,
    ugriz, $JHK$ and Spitzer 3.6 $\mu$m and 4.5 $\mu$m bands compared
    to the models of the star and the debris disk.}
\end{figure}

\section{RESULTS AND DISCUSSION}\label{results}

As discussed in details in \citet{koester09}, the interpretation of
the abundance pattern measured at a white dwarf photosphere is not
always straightforward. Indeed, depending in which of the three phases
of the accretion-diffusion process the star is observed (which we
generally do not know, especially in the case of helium-rich
atmosphere white dwarfs with relatively long diffusion timescales),
the relation between the photospheric abundance and that of the
accreted body differs.

For instance, in the first phase, when the accretion process just
began, the observed relative abundance of each element measured at the
photosphere directly translates into that of the polluting body. Next,
the so-called steady state is reached and the abundance pattern can no
longer be directly interpreted since each element has its own
characteristic diffusion timescale. Fortunately, one can still recover
the relative abundance of the source material by simply applying a
correction that takes into account the difference in diffusion time of
the various elements \citep[see][]{Dupuis}. Finally, when the
accretion ends, each element disappears exponentially with their
corresponding diffusion constant, quickly affecting the abundance
ratios between elements with different timescales. In the case of
J0738+1835, given the strength of the infrared excess, as well as the
large accretion rates needed to explain the photospheric pollution
(see Table 2), one can be confident that this system is currently
accreting and not in the declining phase.

It is generally assumed that systems with important infrared excess
are in the steady state regime, although there is always a small
chance that we are observing this object in the early phase. The
probability of catching an object at that moment depends on the disk
lifetime, which is, however, highly uncertain. Based on simple
assumptions, \citet{jura08} estimates that a median-sized asteroid may
produce a disk that last around $1.5\times 10^5$ yrs, comparable to
the diffusion timescales of the heavy elements in J0738+1835's
photosphere (see below). If dusty disk lifetimes are truly that short,
then the steady state might not be reached and the observed abundance
pattern could be directly interpreted. However, it is also possible
that dusty disks, depending on their viscosity and composition, may
last as long as $10^8$ yrs, like the rings of Saturn. Accordingly,
given our poor understanding of the temporal evolution of
circumstellar disks, caution is well advised when interpreting
observed abundance ratios \citep[for example,
see][]{Rafikov11,Jura12}. The presence of a gaseous component,
however, seem to indicate that J0738+1835's disk will probably evolve
much faster than that. Nevertheless, assuming that a much larger body
such as the one that was tidally disrupted near J0738+1835 could last
longer (a few settling time) than the simple estimates of
\citet{jura08}, then a steady state could be reached.

In any case, the maximum correction applicable on the abundance ratio
of two elements in the steady state regime is only $\sim$1.6 (see
below) for J0738+1835's atmospheric parameters, and our conclusions
are not significantly affected by the exact regime in which J0738+1835
is assumed to be. For simplicity, we will thus simply assume that
J0738+183 is accreting material in the steady state regime for the
rest of our discussion.

\subsection{Convection Zone Models and Diffusion Timescales}

The convection zone models for our present analysis of J0738+1835 have
been built as described in Dufour et al. (2010). In brief, we computed
full stellar structures specified by the values of their surface gravity
and effective temperature (to ease comparison with the spectroscopic
observations), as well as a chemical stratification given by a He-rich
($Z$ = 0.001) outer layer of fractional mass log ($1 - M(r)/M_*$) =
$-$3.0 surrounding a C/O core. Given the relative insensitivity of the
convection zone thickness on the assumed convective efficiency for the
atmospheric parameters appropriate for J0738+1835 (see Dufour et
al. 2010), we only considered the so-called ML2/$\alpha$=1.0 version in
the present paper. Convection zone models were thus computed for five
values of the effective temperature (\Te = 13,000 K to 14,200 K, in
steps of 300 K), and five values of the surface gravity (\logg = 8.1
to 8.9, in steps of 0.2 dex). This domain in parameter space is centered
on the atmospheric parameters for J0738+1835 initially inferred in
Dufour et al. (2010), and more than cover the uncertainty range
ultimately found in the present study. 

The spline coefficients provided by \citet{paquettea} to characterize
the collision integrals involved in diffusion processes have been
widely used in stellar astrophysics. In the case of white dwarfs, they
have first been used by \citet{paquetteb} to estimate the diffusion
timescales of a few representative metals at the base of the
convection zone of sufficiently cool He- and H-rich stars. The exact
same approach has been followed recently by Koester (2009) for
additional elements, including the use of the same rough model of
pressure ionization devised by \citet{paquetteb} for estimating the
average charge of trace heavy elements under white dwarf conditions.
Given the importance of investigating the presence of trace amounts of
metals in cool white dwarfs in the context of the new paradigm of a
planetary origin, our group at Montr\'eal has gone back to the basic
problem of computing still more realistic diffusion coefficients. The
details of this effort will be reported elsewhere, but we point out
here that 1) improvements in the numerical evaluations of the
collision integrals have been implemented, 2) a more physical
description of the screening length has been used, 3) an improved
model of pressure ionization (insuring smoothness of the charge of an
element with depth) has been adopted, and 4) all 27 elements from Li
to Cu in the periodic table have been considered in the calculations.

Figure 16 summarizes our computations of settling timescales at the base
of the convection zone for our He-rich white dwarf models. In the range
of effective temperatures depicted, the diffusion timescale for a given
element increases monotonically with decreasing temperature at fixed
gravity, but the effect is relatively small. On the other hand, in the
range of surface gravity considered in Figure 16, the diffusion
timescale increases relatively quickly with decreasing surface gravity
at fixed temperature. We note that, while the diffusion timescales can
vary significantly with changing atmospheric parameters, the ratios of
timescales between any two elements are much less sensitive. 

\begin{figure}[!ht]
  \plotone{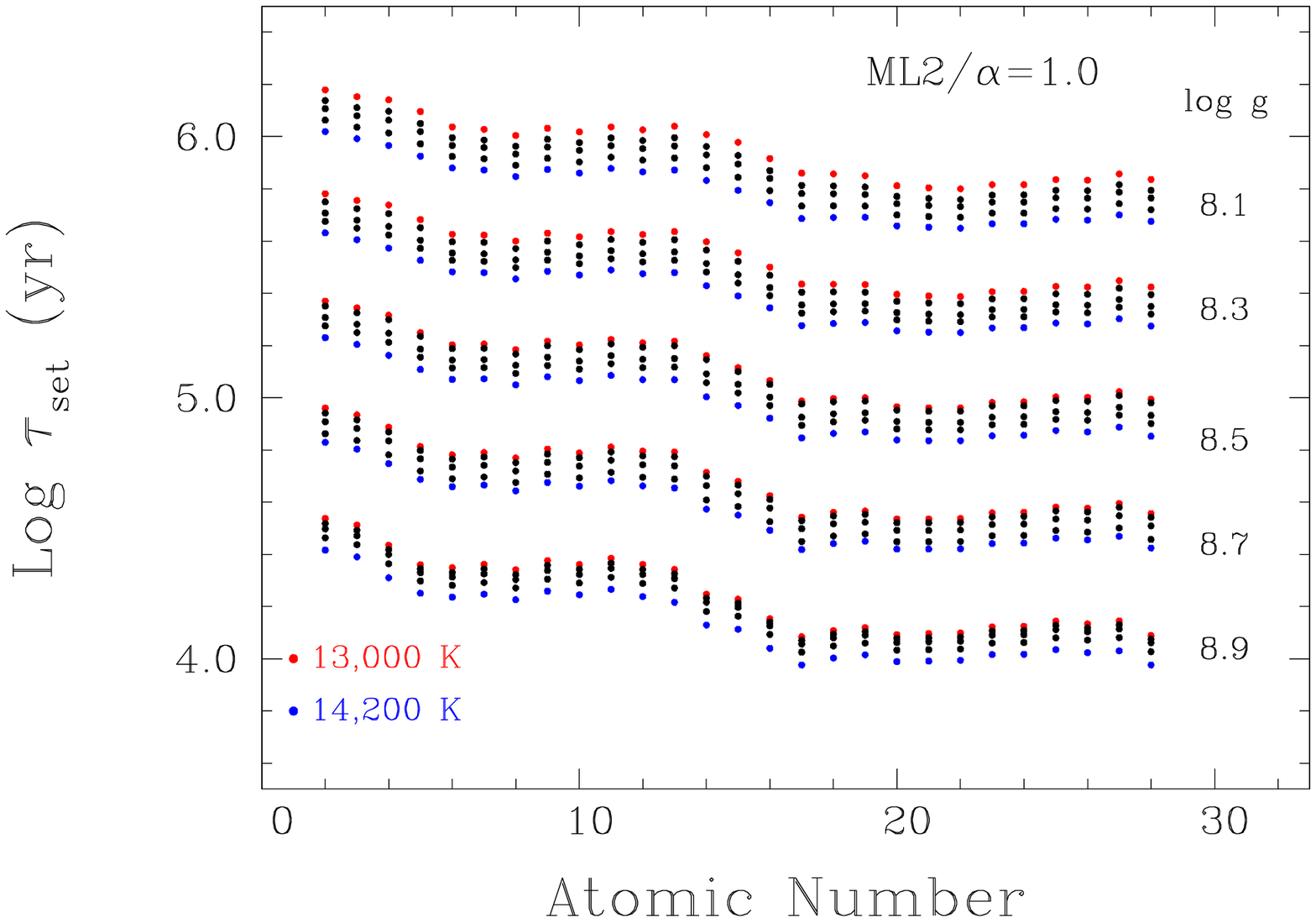} \figcaption[] {Diffusion timescale for all elements
    from Li to Cu (atomic number 3 to 29). Steps of 300 K in effective
    temperature and 0.2 dex in \logg.}
\end{figure}

\subsection{Bulk Composition and Hydrogen}

The derived quantities (abundances, total amount of material found in
the helium convection zone, settling timescales, steady state ratios
and accretion rates) presented in Table 2 reveal a number of
interesting properties about the body that polluted J0738+1835's
photosphere. Not surprisingly, the total amount of heavy elements is
found to be mainly concentrated in the four major constituent of large
rocky bodies like Earth (O, Mg, Si and Fe) with proportions similar to
those found for the most polluted and well studied helium-rich white
dwarfs GD~362, GD~40 and HS2253+8023
\citep{Zuckerman07,Klein2010,Klein2011}.  In total, there is some
$6.6\times10^{23}$ g of accreted material currently present in the
helium convection zone, the largest amount directly detected at the
photosphere of a white dwarf. The object that tidally disrupted near
J0738+1835 was thus at least as massive, and probably even more
massive, than the dwarf planet Ceres. Assuming a typical density of
$\sim$ 2.1 g/cm$^3$, the alleged planet radius was at least 400 km,
and most probably much more considering the unknown amount of material
that i) has already sunk below the convection zone and ii) is still
contained in the surrounding debris disk.

Inspection of Table~2 also reveals a very low amount of hydrogen,
implying that this dwarf planet was very dry, with less than 1\% of
the total mass coming from water ice (note that this is an upper limit
since hydrogen, the lightest element, can only accumulate at the
surface of the white dwarf with cooling age). According to
\citet{jura10}, most of the internal water ice present in a 400 km
object should survive the late stage of evolution. The low amount of
hydrogen found at J0738+1835's photosphere can thus safely be
interpreted as a reflection of the low amount of water ice initially
contained in the accreted object. Hence, the body responsible for
J0738+1835's metal pollution must have formed inside the so-called
"snow line" where the thermodynamic conditions are such that rocky
planetesimals initially possess little to no water \citep[a similar
conclusion was also recently found for a large sample of nearby
helium-rich white dwarfs, see][]{jura12}.

\subsection{Oxygen Budget and Temperature Trends}

Oxygen accounts for more than half of the material accreted onto the
surface of J0738+1835. Another way to strengthen the rocky nature of
the polluting body is to verify that all the accreted oxygen can by
accounted for assuming it is contained in rocky oxygenated minerals
\citep{Klein2010}. In other words, if the oxygen was all contained in
mineral oxides (such as MgO, $\rm Al_2O_3$, $\rm SiO_2$, CaO, $\rm
TiO_2$, $\rm Cr_2O_3$, MnO, FeO, $\rm Fe_2O_3$, and NiO) would the
abundance of all the heavy elements found in the photosphere add up in
sufficient quantity to explain the oxygen abundance? Quantitatively,
this corresponds to \citep[see equation 3 of][]{Klein2010}:
\begin{equation}
  \frac{1}{2}\frac{n(H)}{n(O)} + \sum_{Z}\frac{q(Z)}{p(Z)}\frac{n(Z)}{n(O)} = 1.
\end{equation}
where the heavy element Z is contained in molecule
$Z_{q(Z)}O_{p(Z)}$. Using the values presented in Table 2, we
calculate the steady state ratios and find that the sum adds up only
to around $\sim$ 0.4 depending on how we divide Fe into $\rm Fe_2O_3$
and FeO. Taken at face value, this indicates an oxygen overabundance,
something that is difficult to explain given the severe constraint we
have, from the low amount of hydrogen, on the water content of the
accreted body.

However, as noted above, the oxygen abundance is based only on 3
individual measurements which are quite discrepant between one another
(it is not clear at this point whether this is due to the relatively
lower resolution and signal-to-noise ratio of the observations or
discrepancies in atomic line data/broadening parameters). Since one of
these measurements gives an abundance of log O/He = -4.1 (the other
two give -3.85 and -3.61), which is low enough to reconcile the oxygen
budget, we could be tempted, until more precise measurements become
available, to prefer the low end value in the abundance range
permitted by the uncertainty.

Alternatively, the unusually high oxygen abundance could be explained
by the local thermodynamical condition of the protoplanetery disk at
the distance where the object formed. For instance, since the
equilibrium composition of planetesimal embryos vary with temperature
and pressure at different radii \citep{Bond, Bond2}, one expect, in a
cooler environment, to observe a decrease in the abundances of the
most refractory elements such as Al and C, as well as an increase in
more volatile elements like O \footnote{The C/O ratio is also a very
  important quantity for determining final planetary composition
  \citep{Bond}. Unfortunately, only an upper limit of log C/He $<
  -3.8$ is obtained from the absence of the strongest C
  transitions.}. It is thus possible that most of the oxygen was
actually locked up in more complex molecules than those assumed above
($\rm MgAl_2O_3$, $\rm NaAlSi_3O_8$ for example), which would help to
bring the oxygen balance nearer to 1.

Interestingly, if the accreted object formed from fractionated disk
material whith different thermodynamical conditions than where Earth
formed, then it might also be possible to observe a signature of this
in the form of a correlation in the abundance ratios with condensation
temperatures. Such correlations have been searched for in some heavily
polluted and well studied helium-rich white dwarfs such as GD~40,
PG~1225$-$079 and HS~2253+8023 \citep{Klein2010,Klein2011}.

In the case of GD~40 and PG~1225$-$079, the refractory elements are
substantially enhanced relative to Mg and Earth values.  The detailed
simulations of \citet{Bond} indicate that planets forming in hotter
regions than where Earth formed will be rich in refractories but low
in other elements such as Mg, Si, O and Fe. For the objects analyzed
by \citet{Klein2010,Klein2011}, this is observed for Si, but not for
Mg and Fe, suggesting that other processes could also be
important. For instance, there could have been intense heating in the
red giant phase, silicate vaporization, or even crust removal that
could have affected the final bulk abundances in the bodies that
accreted onto these stars.  Consequently, although the planetary
compositions derived in these stars suggest close formation to the
host star, explaining the observed abundance pattern as a whole still
remains quite challenging.

With 14 heavy elements detected and measured in its photosphere,
J0738+1835 provides us with a unique opportunity to empirically
observe in greater details the abundance pattern for another large
terrestrial-like planet. Following \citet[][see their Figure
18]{Klein2011}, we plot in Figure 17 the observed abundances,
normalized to Mg and bulk Earth, as a function of the condensation
temperatures of each element (taking Fe, instead of Mg, as a reference
produces a very similar plot while using Si shifts the points upward
by about a factor of 1.5). Our measurements suggest that there is
indeed a correlation of the abundance ratios with condensation
temperature, highly refractory elements being depleted (relative to
bulk Earth) while the most volatile elements appear enhanced. We note
that the presence of this trend with condensation temperature depends
little on whether we assume that the accretion occurs in the early
phase or at steady state (in which case a small correction, consisting
of simply multiplying the results by the ratios of the diffusion times
of the considered elements, is applied). The pattern for the most
refractory elements is completely the opposite to what was found for
GD~40 and PG~1225$-$079, clearly indicating a different formation
history and/or subsequent evolution of the body responsible for
J0738+1835's metal pollution.

\begin{figure}[!ht]
  \plotone{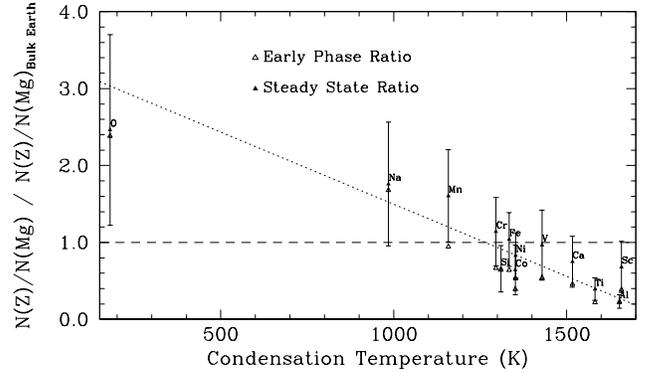} \figcaption[] {Abundances relative to magnesium and
    bulk Earth \citep{Allegre01} as a function of 50\% condensation
    temperature \citep{lodders}. The error bars are plotted only for
    the steady state ratios for clarity (early phase error bars are
    essentially the same). The dotted line is a best least square fit
    to the steady state ratios while the dashed line is to guide the
    eye to Earth-like values.}
\end{figure}

Recent simulations of the formation of terrestrial planets by
\citet{Bond} suggest that objects forming in the inner regions have
high Mg/Si, Al/Si and Ca/Si ratios with a steady transition toward
Earth's values as one moves further out. This trend is a reflection of
the condensation of refractory species in the inner most region and of
Mg-silicate further out. The low abundances that we find for many
refractory species could thus indicate that the planet/asteroid that
polluted J0738+1835 formed further out in cooler regions where the Ca-
and Al-rich inclusions (CAI), as well as other refractory-rich
material, cannot condensate out easily.

Although this simple interpretation is appealing to explain our
results, a few problems remain. For instance, according to
\citet{Bond}'s simulations, while Al is condensing out in the inner
regions of the disk, Na and Al are expected to condense further out
together (species such as $\rm MgAl_2O_3$ and $\rm NaAlSi_3O_8$
condense out over a similar temperature range as Mg silicates). Hence,
it is not clear how it could be possible to deplete Al and Si but not
Na (a similar argument can be applied to Mn as well).

It must be noted, however, that the error bars on some of these
elements are quite large, severely affecting our level of confidence on
the existence of the trend for the low range of condensation
temperature ($\rm T_c < 1200$~K). In particular, Na and Mn abundances
rely on very few measurements and given the signal-to-noise ratio of
our observations in the region where these lines are found (and the
corresponding error bars), it is quite possible that the real
abundances of these elements could lie on the low end range of our
measurements, which would be more compatible with Earth-like
values. Furthermore, it must be noted that the estimation of the
composition of bulk Earth volatile material is quite a difficult task
\citep[see][]{Allegre,Allegre01} and that uncertainties on the Na and
Mn Earth values used for the normalization in Figure 17 are to be
considered as well. For example, taking Na and Mn values from
\citet{Allegre} over those of \citet{Allegre01} would lead to a better
agreement between the observed abundances of these elements and Earth
values.

Since, as we discussed above, the oxygen abundance is also quite
uncertain (the lower abundance of oxygen would also be more compatible
with the oxygen balance argument in certain conditions), another
coherent picture could be that the observed abundances are in fact
close to Earth values for most elements, with only a few exceptions
like Si, Ti and Al, which can confidently be considered depleted
relative to bulk Earth.

Such a pattern, if confirmed with better accuracy, could be
interpreted very differently. For example, it is possible that
differentiation took place in the body that accreted at the surface of
J0738+1835 (in fact, with a minimal radius of at least 400 km,
differentiation most probably occurred). As discussed in
\citet{Klein2010}, elements such as Si, Ca, Ti, Al tend to concentrate
in the crust of a differentiated body. Given that it is these very
same elements that are found to be depleted in our object, the
abundance pattern found in J0738+1835's photosphere could very well be
due to some sort of crust removal. A similar process may explain the
strange abundance pattern found in GALEX J1931+0117 \citep{melis11},
where the planet's outer layers is believed to have been stripped away
by some wind interaction in the AGB phase. Another possibility would
be that a large impact striped away a large portion of the crust,
affecting the final composition that we observe at the white dwarf
photosphere \citep[see][where a similar scenario has been proposed to
explained the peculiar abundances found in NLTT~43806]{Zuckerman11}.

It is unfortunate that a detailed analysis such as this one, even
using data from some of the best observational facilities in the
world, cannot discriminate furthermore between the different possible
scenarios. Although our analysis clearly indicates that the most
refractory elements are depleted relative to Mg and Earth, the case
for a trend with condensation temperature must be considered more
speculative at this point due to the uncertainties in our
measurements. .

Future investigation aiming at refining, among other things, the
oxygen, sodium and manganese abundances, should help to rule out some
of the various scenarios described above. Meanwhile, until better
spectroscopic observations become available, we prefer to remain
cautious and not overinterpret the data. Regardless, our study seems
to indicate that that we are on the verge of entering an exciting new
era of high precision measurements of terrestrial-like planet
compositions from polluted white dwarfs and that appreciable insights
on planetary formation should be accessible in the near future..

\section{CONCLUSION}\label{conclusion}

We presented a detailed analysis, based on high resolution optical
spectroscopic observations, of the DBZ SDSS~J073842.56+183509.06, the
most metal polluted white dwarf currently known. We also obtained IRAC
{\it Spitzer} infrared photometry at 3.6 $\mu$m and 4.5 $\mu$m, which,
combined with earlier $JHK$ photometric measurements, allows us to
better constrain the properties (inner and outer temperature,
inclination) of the bright debris disk surrounding that white dwarf.

We have measured the abundances of 14 different heavy elements and
determined that the total mass of material currently present in the
helium convection zone is $\sim6.6\times10^{23}$ g, indicating that the
object that was tidally disrupted was at least as large as the dwarf
planet Ceres. We also find, from the low amount of hydrogen present in
the photosphere, that the total amount of water ice was less the 1\%
of the mass of this object, suggesting that it formed inside the
so-called snow line.

Our most interesting finding, summarized in Figure 17, is that the
highly refractory species are depleted, relative to Mg and Earth,
while more volatile elements appear to be enhanced, with a possible
correlation with the condensation temperature. Such a trend could be
the signature that the dwarf planet that accreted on J0738+1835 formed
in a lower temperature environment than Earth, where the
thermodynamical conditions are such that the planetesimal embryos'
chemical composition tends to be depleted of more refractory
elements. However, the correlation must be considered uncertain given
the large uncertainties associated with the O, Na and Mn
measurements. Nevertheless, it is clear that the Si/Mg, Ti/Mg and Al/Mg
ratios are lower than bulk Earth values, indicating that the physical
processes that affected the final composition of the dwarf planet were
different than those that occurred for Earth and other polluted white
dwarfs such as GD~40, GD~362 and PG~1225$-$079.

Another way to interpret our findings, if we momentarily ignore the
more uncertain O, Na and Mn abundances, is that differentiation
occurred in the large body that produced the pollution observed on
J0738+1835. With a radius of at least 400 km, a crust containing
preferentially Si, Ti, Al and Ca could have formed and been removed,
potentially explaining part of the abundance pattern that we find.

Improvements, allowing us to discriminate between these various
scenarios, albeit difficult, should be obtainable from longer
observing runs on large telescopes (the combination of high resolution
with high signal-to-noise ratio is the key to obtain precise
results). This work, as well a many other recent studies \citep[for
example,][]{Zuckerman07, Zuckerman10, melis11, Klein2010, Klein2011}
demonstrate, once more, the extraordinary potential of metal
contaminated white dwarfs to study rocky extrasolar planet/asteroid
bulk compositions. These objects are unique probes that can serve as
perfect testbeds for planetary formation and evolution theories,
providing information that cannot be obtained from any other way. As
our understanding of the physical processes at play increases, as well
as our confidence in the interpretation of the chemical patterns found
in these stars, reliable bulk composition of rocky extrasolar objects,
conceivably with a level of precision comparable, if not better, to
what is possible for similar objects in the solar system, should soon
be provided from metal polluted white dwarf studies.

\acknowledgements We would like to thank Adam Burgasser for sharing
some of his telescope time and Jade C. Bond for many useful
discussions. This work was supported in part by NSERC Canada and FQRNT
Qu\'ebec. P.D is a CRAQ postdoctoral fellow. G.F. acknowledges the
contribution of the Canada Research Chair Program. We acknowledge the
use of the VALD Database
\citep{Piskunov,Ryabchikova,Kupka99,Kupka00}. Keck telescope time was
granted by NOAO, through the Telescope System Instrumentation Program
(TSIP). TSIP is funded by NSF.  The authors wish to recognize and
acknowledge the very significant cultural role and reverence that the
summit of Mauna Kea has always had within the indigenous Hawaiian
community. We are most fortunate to have the opportunity to conduct
observations from this mountain.

\end{document}